\documentclass[conference, 9pt]{IEEEtran}
\IEEEoverridecommandlockouts

\usepackage{cite}
\usepackage{amsmath,amssymb,amsfonts}
\usepackage[ruled]{algorithm}
\usepackage{algorithmic}
\usepackage{graphicx}
\usepackage{flafter}
\usepackage{textcomp}
\usepackage{xcolor}
\usepackage{booktabs,tabularx,multirow,threeparttable}
\usepackage{colortbl}
\usepackage{siunitx}
\definecolor{TergelBlue}{RGB}{0,83,155}

\newif\ifshowtergelcomments
\showtergelcommentstrue

\def\BibTeX{{\rm B\kern-.05em{\sc i\kern-.025em b}\kern-.08em
    T\kern-.1667em\lower.7ex\hbox{E}\kern-.125emX}}
\begin{document}

\bstctlcite{IEEEexample:BSTcontrol}

\title{RE-CAM: Readout-Enhanced Content-Addressable Memory for Quantitative Similarity Search}


\author{
  \IEEEauthorblockN{
    Yi-Chun Liao\textsuperscript{1},
    Dongyoung Lee\textsuperscript{1, 2},
    Jiazheng Chen\textsuperscript{1},
    Tergel Molom-Ochir\textsuperscript{1},
    Tomohisa Kawakami\textsuperscript{1}, \\
    Tania Roy\textsuperscript{1},
    Ik-Joon Chang\textsuperscript{2},
    X. Sharon Hu\textsuperscript{3},
    Hai ``Helen" Li\textsuperscript{1},
    Yiran Chen\textsuperscript{1}
  }
  \IEEEauthorblockA{
    \textsuperscript{1}Duke University, Durham, NC, USA\\
    \textsuperscript{2}Kyung Hee University, Seoul, South Korea \\
    \textsuperscript{3}University of Notre Dame, Notre Dame, IN, USA\\
    \{yi-chun.liao, yiran.chen\}@duke.edu
  }
}

\maketitle

\begin{abstract}
    Similarity search is essential to many applications but incurs massive data movement when executed on conventional architectures. Content-addressable memory (CAM) is a promising hardware substrate, as it allows similarity search directly in the CAM array, minimizing data movement and offering both low latency and high energy efficiency. However, conventional CAM array readout produces only binary match results, discarding quantitative similarity information that could be further exploited. Extracting this information requires interpreting the CAM array output voltage that evolves during a CAM search, making the sensing timestamp a crucial design variable alongside the reference voltage and decoding parameter. In this work, we propose RE-CAM, a readout enhancement framework for CAM that is agnostic to memory technology and application, enabling CAM array readout to approximate linear and nonlinear functions with one CAM search. By reusing the comparators to sense the time-varying output voltage multiple times, jointly optimizing the three design variables under system constraints and hardware nonidealities, and co-designing the optimization algorithm and reconfigurable hardware, RE-CAM achieves accurate approximation with low energy and area overhead without introducing a readout throughput bottleneck. Across linear, piecewise-linear, sigmoid, and exponential functions, RE-CAM reduces normalized root mean square error by 10.89\% to 24.34\% for general approximation and 29.43\% to 46.73\% when exploiting dataset distribution, compared to a shared uniform-reference ADC baseline with the same comparison budget. Relative to the same baseline, RE-CAM also achieves $8\times$ higher search throughput while reducing energy per search and area by 47\% and 39\%, respectively.
\end{abstract}

\begin{figure}[!t]
    \centering
    \includegraphics[width=\columnwidth]{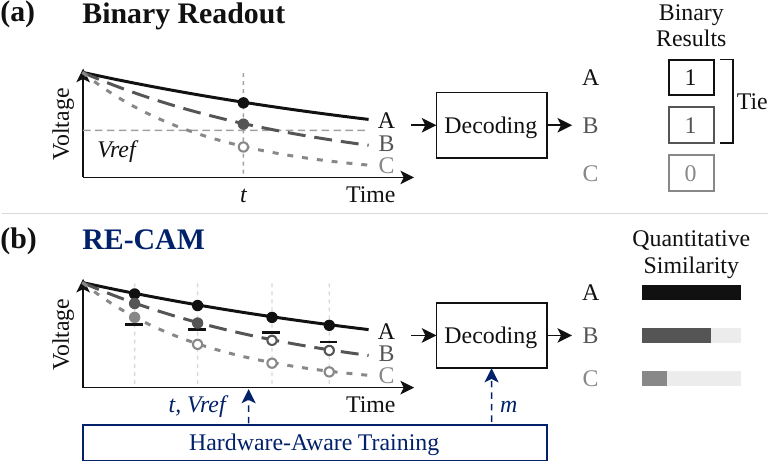}
    \caption{Conceptual readout enhancement. (a) Binary readout merges different similarity levels. (b) RE-CAM jointly optimizes sensing timestamps, reference voltages, and decoding parameters to enable quantitative similarity readout with one CAM search (i.e., one precharge-discharge phase).}
    \label{fig:teaser}
\end{figure}

\section{introduction} \label{sec:introduction}

Similarity search, which compares a query vector with candidate vectors using measures such as Hamming distance and dot product to select the most similar candidates, is fundamental to many modern applications such as recommender systems \cite{RecommenderSystem}, retrieval-augmented generation \cite{RAG}, and DNA classification \cite{DNAClassification}. However, processing large datasets on conventional von-Neumann architectures incurs substantial data movement overhead. Previous work \cite{offchip} reports that for data-intensive applications, data transfer and off-chip memory access can account for up to 90\% of the overall latency and energy.

As a promising solution to the architectural limitations, in-memory search (IMS) has attracted increasing attention in recent years. Content-addressable memory (CAM) \cite{CAMSurvey} realizes IMS through specialized cell and peripheral circuitry, providing computation tailored to similarity search while avoiding the transfer of stored candidate vectors. CAM compares a query vector with all candidate vectors stored in a physical array in parallel, and the result readout comprises two stages: \textit{sensing}, which converts the analog array output voltage into digital observations, and \textit{decoding}, which maps these observations to final search results according to a predefined decoding scheme. While these capabilities make CAM well-suited for search-based data-intensive applications, conventional designs support only binary match/mismatch readout, discarding the quantitative similarity information contained in the output voltage.

The limited expressiveness of CAM array readout restricts the performance of existing CAM-based applications and may also limit CAM's broader applicability. For instance, classification using hyperdimensional computing \cite{HDCIntro} involves identifying the best match among vectors with high dimensionality, which a single CAM array fails to hold. Previous work \cite{HDC} partitions the high-dimensional candidate vectors into multiple subvectors that can fit into separate CAM arrays, identifies the local best match within each array, and selects the final match result by majority voting. Nevertheless, local binary best-match results discard the quantitative similarity information for both winning and nonwinning candidates, causing different local similarity levels to yield identical contributions to the global vote count. This information loss is reported to decrease classification accuracy on ISOLET \cite{ISOLET} from 92.5\% under ideal unpartitioned search to 44.0\% under partitioned search with majority voting \cite{HDC}. Approximating Hamming distance and summing the partitioned results instead could reduce information loss and bring partitioned search closer to the accuracy of unpartitioned search. Besides, one recent work CAMformer \cite{CAMformer} explores the potential of leveraging CAM arrays for attention computation, but relies on costly analog-to-digital converters (ADCs) for readout and sophisticated digital circuitry for activation. A readout design that allows direct approximation from output voltage to activation functions, such as ReLU, sigmoid, and exponential (for softmax), is therefore highly desirable.

\begin{figure*}[!t]
    \centering
    \includegraphics[width=\linewidth]{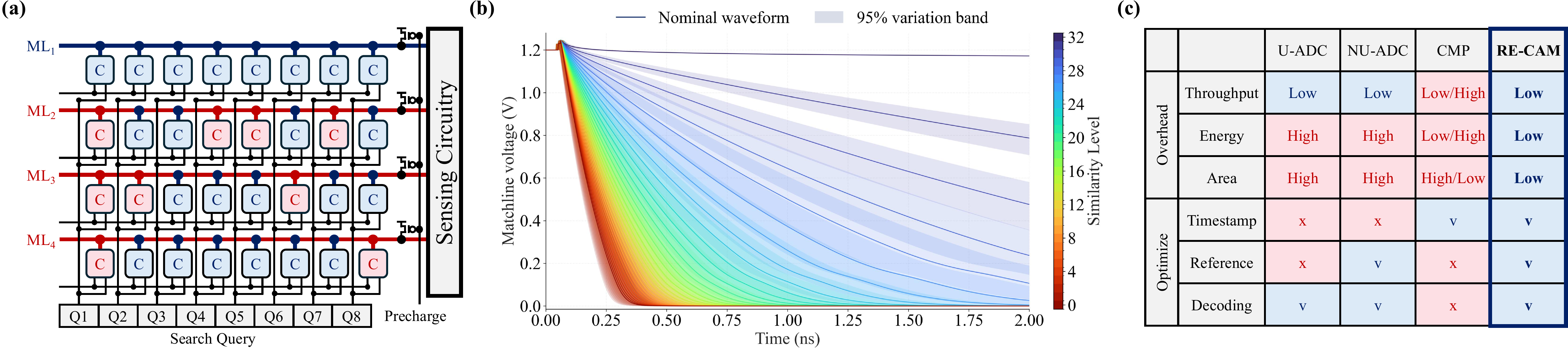}
    \caption{(a) Generic CAM array architecture. (b) SPICE-simulated matchline voltage waveforms across similarity levels (32 - Hamming distance) of a $1\times32$ binary CAM array with 10\% discharge speed variation. Nominal waveforms assume no variation. (c) Comparison of existing quantitative readout approaches by throughput, energy, and area overhead and optimization of the three design variables, assuming no sensing-circuit sharing across matchlines.}
    \label{fig:background}
\end{figure*}

However, the time-varying output voltage of a CAM array complicates function approximation. Specifically, different similarity levels are best separated with different sensing timestamps and reference voltages. Moreover, approximating different functions prioritizes the separation of different similarity levels. Realizing quantitative CAM array readout therefore requires answering \textit{\textbf{when to sense, which reference to compare with, and how to decode the binary observations}}, three questions that, to the best of our knowledge, have not yet been jointly considered for quantitative CAM array readout.

In this paper, we introduce RE-CAM, a readout enhancement framework for CAM that is agnostic to memory technology and application, extending CAM array readout from binary match results to approximations of linear and nonlinear functions. RE-CAM reuses comparators (CMPs) to sense the time-varying output voltage multiple times, and jointly optimizes sensing timestamps, reference voltages, and decoding parameters under system constraints and hardware nonidealities. RE-CAM co-designs the optimization algorithm and reconfigurable hardware to implement dedicated encoding for binary observations, reducing energy and area overhead without introducing a readout throughput bottleneck. Fig.~\ref{fig:teaser} presents the high-level concept of RE-CAM. The main contributions are as follows:

\begin{itemize}
    \item \textbf{Algorithm:} A hardware-aware training algorithm that optimizes the readout schedule, comprising sensing timestamps, reference voltages, and decoding parameters, to approximate a target function under system constraints and hardware nonidealities.
    \item \textbf{Hardware:} A CAM macro that executes the readout schedule through reconfigurable sensing and decoding, enabling the same hardware to approximate different functions without requiring costly ADCs or introducing a readout throughput bottleneck.
    \item \textbf{Dataset-Specific Enhancement:} An enhancement technique that exploits the dataset distribution to prioritize the approximation accuracy of frequently encountered similarity levels.
\end{itemize}

We evaluate RE-CAM on a 32$\times$32 binary CAM array using four target functions: linear, piecewise-linear, sigmoid, and exponential. Compared to a shared uniform-reference ADC baseline with the same comparison budget (i.e., the maximum number of comparisons per candidate vector per CAM search), RE-CAM reduces normalized root mean square error (NRMSE) by 10.89\% to 24.34\% for general approximation and 29.43\% to 46.73\% when exploiting dataset distribution. RE-CAM also achieves $8\times$ higher search throughput while reducing energy per search and area by 47\% and 39\%, respectively.

\section{background and motivation}

\subsection{Content-Addressable Memory}

Whereas conventional memory retrieves data by address, content-addressable memory (CAM) \cite{CAMSurvey} searches stored data for entries that match a query and returns their addresses. Fig.~\ref{fig:background}(a) presents a generic CAM array architecture. A CAM cell (C) holds one dimension of the data, with each column associated with a searchline (SL) representing a respective dimension, and every row connected to a matchline (ML) representing a multidimensional candidate vector. During a CAM search, each ML is first precharged, and each dimension of the query vector (Q1 to Q8) is then applied to the corresponding SL. All cells compare applied and stored values in parallel, with each mismatched cell (red) enabling a discharge path on its ML. This work uses a binary CAM array, where each cell stores one bit, and the number of mismatched cells in an entry equals the Hamming distance between the query and the stored candidate vector. We refer to each possible number of matched bits as a \textit{similarity level}, with higher levels indicating greater similarity. Fig.~\ref{fig:background}(b) shows SPICE-simulated ML voltage waveforms for different similarity levels of a binary $1\times32$ CAM array with 10\% discharge speed variation (detailed in Section~\ref{sec:nonidealities}). Ideally, the ML of a fully matched entry remains at the precharge voltage, whereas MLs of lower similarity levels discharge more rapidly. The ML voltage thus reflects the similarity between the query and each candidate vector.

Depending on the design of the cell and readout circuitry, CAM supports various cell data representations (e.g., binary \cite{CAMformer}, ternary \cite{TAPCAM}, multi-bit \cite{HDC}, and analog \cite{Analog}) and decoding schemes (e.g., binary matching \cite{NNSinMCAM, TAPCAM} and quantitative similarity \cite{CAMformer}). The parallel IMS capability of CAM enhances the efficiency of diverse workloads, including tree-based machine learning \cite{RETENTION}, hyperdimensional computing \cite{HDC}, Monte Carlo tree search\cite{MCTS}, memory-augmented neural network \cite{CAMMANN}, and memory-based deep reinforcement learning \cite{CAMDRL}. 

Despite supporting quantitative readout, existing implementations are designed only for Hamming-distance recovery \cite{ExploringHDC, DUAL, CAMformer}, and incur substantial overhead in search throughput, energy, or area, severely limiting their applicability. These limitations motivate an efficient and reconfigurable CAM array readout design that supports the approximation of different functions beyond Hamming distance.

\subsection{Existing Quantitative Readout Approaches}

In this section we examine the overhead and limitations of existing quantitative readout approaches, as summarized in Fig.~\ref{fig:background}(c).

\textbf{Uniform-Reference ADC (U-ADC):} An \textit{n}-bit U-ADC uniformly quantizes the ML voltage range into $2^n$ bins at a selected timestamp. Although this approach supports function approximation with a single CAM search \cite{CAMformer}, uniform voltage bins may fail to separate similarity levels that are important for approximating the target function. Higher-resolution ADCs may improve separation but incur substantial area overhead and can dominate energy consumption \cite{RAELLA}.

\textbf{Nonuniform-Reference ADC (NU-ADC):} An NU-ADC allows nonuniform quantization of the ML voltage range. For function approximation, previous works \cite{NL-DPE, RACE-IT, NL-ADC} generally use this flexibility to uniformly quantize the function’s output range. NL-DPE \cite{NL-DPE} and RACE-IT \cite{RACE-IT} implement similar quantization using analog CAM arrays for crossbar array readout. However, analog CAM remains an emerging technology, with challenges in precision, reliability, and scalable integration limiting its readiness for widespread deployment \cite{CAMPerspective}. Alternatively, NL-ADC \cite{NL-ADC} realizes nonlinear conversion using a ramp ADC with comparators and time-to-digital converters (TDCs). Nevertheless, its reference is generated by the associated crossbar array, limiting its applicability to other architectures. Moreover, even programmable references may fail to separate similarity levels whose ML voltages are too close at the selected timestamp.

\textbf{CMP:} A CMP produces a binary observation indicating whether the ML voltage exceeds the reference voltage at a selected timestamp. To enable expressive readout, multiple comparisons are required. Previous work \cite{ExploringHDC} uses as many CMPs as CAM cells per ML to recover the exact Hamming distance, introducing significant area overhead. DUAL \cite{DUAL} reuses one CMP per ML but activates at most 7 dimensions per CAM search. Consequently, processing a 64-dimensional candidate vector requires at least 10 CAM searches, consuming excessive energy and degrading search throughput. Furthermore, its frequent sensing requires 10 GHz timing control. Beyond these limitations, both designs focus on Hamming-distance recovery without exploring the approximation of other functions.

\subsection{Motivation} \label{sec:motivation}

Function approximation with CAM array readout depends on three coupled design variables: sensing timestamp, reference voltage, and decoding parameter. The sensing timestamp and reference voltage jointly determine how each comparison separates similarity levels into two groups, while the decoding parameters determine how the resulting binary observations are mapped to a numerical output. As shown in Fig.~\ref{fig:background}(b), the voltage differences between adjacent similarity levels vary over time, affecting which levels can be separated at each sensing timestamp. Hardware nonidealities further affect these voltage differences through variation in ML discharge behavior. Meanwhile, the target function determines which similarity levels need to be separated and how the resulting observations should be mapped to a numerical output. These dependencies motivate jointly optimizing the three design variables under system constraints, such as CAM search frequency, system clock rate, and minimum sensing interval, and hardware nonidealities, such as ML discharge speed variation, comparator residual offset, and comparator sensing noise.

RE-CAM reuses one CMP per ML to collect multiple observations per CAM search. By co-designing algorithm and hardware, RE-CAM enables accurate function approximation with low energy and area overhead without introducing a readout throughput
bottleneck.

\section{re-cam framework}

This section presents RE-CAM, a readout enhancement framework for CAM that is agnostic to memory technology and application. We first formulate the readout optimization problem, considering system constraints and hardware nonidealities, and propose a hardware-aware training algorithm to optimize the readout schedule. We then present a co-designed CAM macro to execute the schedule through reconfigurable sensing and decoding. Finally, we provide a dataset-specific enhancement that prioritizes approximation accuracy on the target dataset over the average accuracy across the full input range.

\subsection{Hardware-Aware Training} \label{sec:alg}

\subsubsection{Problem Formulation}

RE-CAM approximates a target function by repeatedly sensing the ML and mapping observations to a numerical output. For a $d$-dimensional binary CAM array, let $x\in\{0,\ldots,d\}$ denote the similarity level, and $f(x)$ the target value. Let $T$ denote the available system clock ticks within one CAM search, and $\mathcal{T}$ all chronologically ordered combinations from $T$ with a minimum sensing interval of $\tau_{\mathrm{cmp}}$ (i.e., comparator latency) between consecutive timestamps. Sub-clock-tick sensing requires more stringent timing control and is outside the scope of this paper. For each candidate $\mathbf{t}=[t_0,\ldots,t_{S-1}]\in\mathcal{T}$, $S=|\mathbf{t}|$ denotes the number of sensing timestamps. The corresponding reference voltages $\mathbf{r}=[r_0,\ldots,r_{S-1}]$ and decoding parameters $\mathbf{m}=[m_0,\ldots,m_{S-1},m_S]$ are jointly optimized.

Each comparison at $t_i$ produces a binary observation indicating whether the ML voltage exceeds $r_i$. Given that lower and higher similarity levels are more separable earlier and later, respectively, we constrain the reference voltages to separate levels from low to high, producing nominal observation sequences (i.e., without nonidealities) of the form $1\cdots10\cdots0$, termed \textit{first-zero encoding}. We then map each sequence to one of $S+1$ categories: predicting $\widehat{f}(x)=m_k$ if the first zero occurs at $t_k$, or $\widehat{f}(x)=m_S$ if all observations are one.

We jointly optimize $\mathbf{t}\in\mathcal{T}$, $\mathbf{r}\in\mathcal{R}(\mathbf{t})$, and $\mathbf{m}$ to minimize the expected mean squared error (MSE):
\begin{equation}
\mathcal{L}(\mathbf{t},\mathbf{r},\mathbf{m})
=
\frac{1}{d+1}\sum_{x=0}^{d}
\mathbb{E}_{\xi}
\left[
\bigl(
\widehat{f}(x;\mathbf{t},\mathbf{r},\mathbf{m},\xi)-f(x)
\bigr)^2
\right],
\label{eq:loss}
\end{equation}
where $\xi$ captures discharge and comparator variations, and $\mathcal{R}(\mathbf{t})$ enforces valid reference-voltage ranges and nominal sequence ordering.

\subsubsection{Hardware Nonideality Modeling} \label{sec:nonidealities}

To evaluate $\mathcal{L}(\mathbf{t},\mathbf{r},\mathbf{m})$, we model the noisy search with three hardware nonidealities: ML discharge speed variation, comparator residual offset, and comparator sensing noise. ML discharge speed variation is modeled by scaling the nominal ML voltage waveforms:
\begin{equation}
\widetilde{V}_x(t)=V_x(t/a),
\label{eq:discharge_variation}
\end{equation}
where $V_x(t)$ denotes the nominal ML voltage of similarity level $x$ at time $t$, and $\widetilde{V}_x(t)$ denotes the corresponding voltage under discharge speed variation. The random time-scaling factor $a$ follows a lognormal distribution with unit mean and standard deviation $\rho$. This distribution ensures a positive time-scaling factor and is approximately Gaussian for small $\rho$. We independently sample one factor per ML per CAM search and hold it fixed across all sensing timestamps within that search. Fig.~\ref{fig:background}(b) presents the ML discharge speed variation modeled with $\rho = 10\%$.

Consistent with the measured comparator offset and sensing noise characteristics \cite{CmpNoise}, we model the residual offset and sensing noise using Gaussian distribution with zero mean and standard deviations $\sigma_{\mathrm{off}}$ and $\sigma_{\mathrm{n}}$. The residual offset is held constant across comparisons within each CAM search, while sensing noise is sampled independently for each comparison.

\begingroup
\color{black}
\begin{algorithm}[!t]
\caption{Hardware-Aware Readout Training}
\label{alg:readout_training}
\footnotesize
\raggedright
\renewcommand{\algorithmicrequire}{\textbf{Input:}}
\renewcommand{\algorithmicensure}{\textbf{Output:}}
\begin{algorithmic}[1]
\REQUIRE ML waveforms $V$, $f$, $T$, $\tau_{\mathrm{cmp}}$,
$\rho$, $\sigma_{\mathrm{off}}$, $\sigma_{\mathrm{n}}$,
$\mathcal{R}$, \textit{epoch}
\ENSURE $\mathbf{t}$, $\mathbf{r}$, $\mathbf{m}$
\STATE $\mathcal{T}\gets
\operatorname{GenerateFeasibleCombinations}(T,\tau_{\mathrm{cmp}})$
\STATE $\mathit{bestLoss}\gets\infty$
\FOR{$\mathbf{t}\in\mathcal{T}$}
    \STATE $S\gets|\mathbf{t}|$
    \STATE $\mathbf{r}\gets
    \operatorname{Initialize}(\mathcal{R}(\mathbf{t}))$
    \FOR{\textit{i} in range(\textit{epoch})}
        \STATE $\mathbf{P}\gets
        \operatorname{ComputeCategoryProbability}(V,\mathbf{t},
        \mathbf{r},\rho,\sigma_{\mathrm{off}},\sigma_{\mathrm{n}})$
        \STATE $\mathbf{m}\gets
        \operatorname{ClosedFormSolution}(\mathbf{P},f)$
        \STATE $\mathit{loss}\gets
        \mathcal{L}(\mathbf{t},\mathbf{r},\mathbf{m})$
        \STATE $\mathit{gradient}\gets
        \operatorname{ComputeGradient}(\mathbf{t},
        \mathbf{r},\mathbf{m})$
        \STATE $\mathbf{r}\gets
        \operatorname{SLSQP}(\mathbf{r},\mathit{loss},
        \mathit{gradient},\mathcal{R}(\mathbf{t}))$
        \STATE \textbf{break if} $\operatorname{Converged}()$
    \ENDFOR
    \STATE $\mathbf{P}\gets
    \operatorname{ComputeCategoryProbability}(V,\mathbf{t},
    \mathbf{r},\rho,\sigma_{\mathrm{off}},\sigma_{\mathrm{n}})$
    \STATE $\mathbf{m}\gets
    \operatorname{ClosedFormSolution}(\mathbf{P},f)$
    \STATE $\mathit{loss}\gets
    \mathcal{L}(\mathbf{t},\mathbf{r},\mathbf{m})$
    \IF{$\mathit{loss}<\mathit{bestLoss}$}
        \STATE $\mathit{bestLoss}\gets\mathit{loss}$
        \STATE $\mathit{best}\gets
        (\mathbf{t},\mathbf{r},\mathbf{m})$
    \ENDIF
\ENDFOR
\STATE $(\mathbf{t},\mathbf{r},\mathbf{m})\gets\mathit{best}$
\RETURN $\mathbf{t},\mathbf{r},\mathbf{m}$
\end{algorithmic}
\end{algorithm}
\endgroup

\subsubsection{Training Algorithm}

With the nonideality models established, we evaluate $\mathcal{L}(\mathbf{t},\mathbf{r},\mathbf{m})$ with the probabilities of decoding categories. Let $\delta$ denote the residual offset. Assuming that ML discharge speed variation, comparator residual offset, and comparator sensing noise are mutually independent, the probability that comparison $i$ outputs one, conditioned on $a$ and $\delta$, is:
\begin{equation}
\begin{aligned}
p_i(x\mid a,\delta)
=\Phi(\frac{\widetilde{V}_x(t_i)-(r_i+\delta)}
{\sigma_{\mathrm{n}}})
\end{aligned}
\label{eq:comparison_probability}
\end{equation}
where $\Phi$ is the standard Gaussian cumulative distribution function (CDF). Since sensing noise is independent across comparisons, the probability of each decoding category is:
\begin{equation}
P_k(x;\mathbf{t},\mathbf{r})=
\begin{cases}
\displaystyle
\mathbb{E}_{a,\delta}
\left[
\left(\prod_{i=0}^{k-1}p_i\right)(1-p_k)
\right],
&0\le k < S,\\[6pt]
\displaystyle
\mathbb{E}_{a,\delta}
\left[\prod_{i=0}^{S-1}p_i\right],
&k=S,
\end{cases}
\label{eq:category_probability}
\end{equation}
where category $k < S$ corresponds to the first zero occurring at $t_k$, while category $S$ represents an all-ones sequence. The expected MSE can therefore be expressed as the probability-weighted error across decoding categories:
\begin{equation}
\mathcal{L}(\mathbf{t},\mathbf{r},\mathbf{m})
=
\frac{1}{d+1}
\sum_{x=0}^{d}\sum_{k=0}^{S}
P_k(x;\mathbf{t},\mathbf{r})\bigl[m_k-f(x)\bigr]^2.
\label{eq:probabilistic_loss}
\end{equation}

For fixed $\mathbf{t}$ and $\mathbf{r}$, the closed-form solution for each $m_k$ is the average of $f(x)$ weighted by $P_k(x;\mathbf{t},\mathbf{r})$. We therefore solve $\mathbf{m}$ analytically and optimize $\mathbf{r}$ using gradients, as the Gaussian CDF $\Phi$ makes the expected loss differentiable with respect to $\mathbf{r}$. 

The overall hardware-aware training is summarized in Algorithm~\ref{alg:readout_training}. As shown in the algorithm, the outer loop enumerates every timestamp combination $\mathbf{t}\in\mathcal{T}$. For each combination, the inner loop uses SLSQP \cite{SLSQP, Scipy} to optimize $\mathbf{r}$ subject to $\mathcal{R}(\mathbf{t})$. At each iteration, we compute the category probabilities, solve $\mathbf{m}$ analytically, compute the loss and gradients, and update $\mathbf{r}$. After training all combinations, we select the readout schedule with the minimum loss. Measured noisy ML waveforms can directly replace the discharge speed variation model without changing the optimization procedure.

\begin{figure}[!t]
    \centering
    \includegraphics[width=\columnwidth]{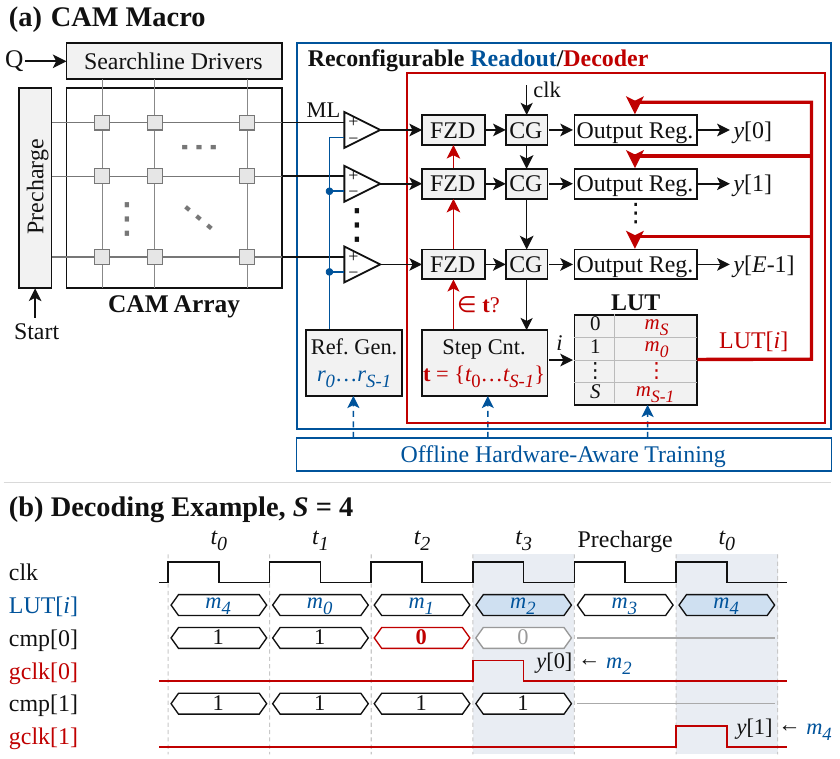}
    \caption{Proposed RE-CAM macro (\textit{E} output registers, \textit{S} sensing), where $y[e]$ denotes the output for entry $e$. (a) Comparators sense matchlines at clock tick $t_i$ with reference $r_i$ during one CAM search. A shared step counter selects entry $i$ of the $(S+1)$-entry LUT, which is broadcast to all output registers. For each ML, the first-zero detector (FZD) enables the clock gate (CG) once, enabling the output register to store $m_i$ one cycle after the first zero at $t_i$, or $m_{S}$ at $t_0$ of the next CAM search if no zero is observed. (b) Decoding example with $S=4$: entry 0 senses $1100$ and stores $m_2$ at $t_3$, ignoring the later zero; entry 1 senses $1111$ and stores $m_4$ at $t_0$ of the next CAM search.}
    \label{fig:architecture}
\end{figure}

\subsection{CAM Macro Architecture}

To execute the optimized readout schedule efficiently, we design hardware that supports reconfigurable sensing and decoding without introducing a throughput bottleneck. Fig.~\ref{fig:architecture}(a) shows the proposed CAM macro, which reuses one CMP per ML for sensing and employs lookup-table-based (LUT-based) decoding. Sensing and decoding overlap to retain the CAM array's native search throughput. After offline hardware-aware training, $\mathbf{t}$ is programmed into the step counter, $\mathbf{r}$ into the reference generator, and $\mathbf{m}$ into the LUT.

During each CAM search, the ML voltages are compared with $r_i$ at sensing timestamp $t_i$. A shared step counter sequentially accesses the LUT and broadcasts the corresponding $m_i$ to all output registers with a one-clock-cycle delay relative to sensing. For each ML, a first-zero detector (FZD) identifies the first zero and enables the associated clock gate (CG), allowing its output register to capture $m_i$ in the following clock cycle. Multiple output registers can therefore be updated in parallel using a single LUT access. Once updated, each register retains its value, and subsequent observations for that ML are ignored. Fig.~\ref{fig:architecture}(b) illustrates this behavior: the sequence $1100$ has its first zero at $t_2$ and produces $m_2$, whereas $1111$ receives $m_4$ at $t_0$ of the next CAM search. LUT access and first-zero detection proceed in parallel, enabling incremental decoding during the search. The final $m_S$ update overlaps the beginning of the next CAM search and completes before its first output update, allowing the readout to retain the CAM array's native search throughput.

First-zero encoding is central to efficient lookup and decoding. By exploiting ML discharge characteristics and explicitly constraining nominal sequence ordering, we restrict nominal observation sequences to the form $1\cdots10\cdots0$. Compared with directly indexing the full $S$-bit observation sequence, first-zero encoding reduces the number of LUT entries from $2^S$ to $S+1$, corresponding to the $S$ possible first-zero positions and the all-ones case. The sensing-step index directly identifies the LUT entry for a newly detected first zero, eliminating the need to buffer the complete observation sequence or perform separate per-ML lookups. A per-ML flag tracks whether a result has already been captured and controls whether to store the shared LUT value, allowing multiple output registers to update in parallel. This enables decoding across all MLs with a single sequentially accessed LUT and minimal per-ML control state.

\subsection{Dataset-Specific Enhancement}

The similarity level between query and candidate vectors generally follows a nonuniform distribution. For instance, Hamming distance between random high-dimensional binary vectors exhibits a bell-shaped distribution. RE-CAM exploits such distributions through prioritizing the separation of frequently encountered similarity levels.

By calibrating the similarity distribution of a given dataset, RE-CAM estimates $q(x)$, the probability of encountering similarity level $x$, from representative query-candidate pairs. Replacing the uniform weighting in Eq.~\eqref{eq:probabilistic_loss} with $q(x)$ gives the dataset-specific objective:
\begin{equation}
\mathcal{L}_{\mathrm{cal}}(\mathbf{t},\mathbf{r},\mathbf{m})
=
\sum_{x=0}^{d}q(x)
\sum_{k=0}^{S}
P_k(x;\mathbf{t},\mathbf{r})
\bigl[m_k-f(x)\bigr]^2,
\label{eq:calibrated_loss}
\end{equation}
where $q(x)\geq 0$ and $\sum_{x=0}^{d}q(x)=1$.

The training procedure remains unchanged. For fixed $\mathbf{t}$ and $\mathbf{r}$, each decoding parameter $m_k$ is solved analytically as the average of $f(x)$ weighted by $q(x)P_k(x;\mathbf{t},\mathbf{r})$. Reference optimization and timestamp selection then use $\mathcal{L}_{\mathrm{cal}}$, adapting the readout schedule to minimize the expected approximation error under the specific distribution.

\begin{table}[!t]
\centering
\caption{evaluation configuration}
\label{tab:setup}
\footnotesize
\setlength{\tabcolsep}{4pt}
\renewcommand{\arraystretch}{1.08}
\begin{tabularx}{\columnwidth}{@{}l>{\raggedright\arraybackslash}X@{}}
\toprule
\multicolumn{2}{@{}l}{\textit{Hardware Implementation}} \\
\midrule
CAM Array & 32$\times$32 array (10T CMOS binary cells); SL 0.6 V; 65 nm\\
Sensing & 4 shared 3-bit U-ADCs; 500-MS/s\\
& 4 shared 3-bit NU-ADCs; 500-MS/s\\
& 32 CMPs; 4 GHz\\
Decoding & Reconfigurable decoder with 32 output registers;\\
& 1 8-entry FP16 LUT; 1.0 V; 65 nm\\
\midrule
\multicolumn{2}{@{}l}{\textit{System Constraints and Hardware Nonidealities}} \\
\midrule
Constraints & 500 MHz CAM search; 4 GHz clock; $\tau_{\mathrm{cmp}} = 70$ ps\\
Nonidealities & $\rho=10\%$; $\sigma_{\mathrm{off}} = 1$ mV \cite{CmpNoise}; $\sigma_{\mathrm{n}} = 1$ mV \cite{CmpNoise} \\
\midrule
\multicolumn{2}{@{}l}{\textit{Target Functions ($x\in\{0,\ldots,32\}$)}} \\
\midrule
Linear & $f(x)=x$ \\
Piecewise-Linear & $f(x) = 
\begin{cases}
0.5(x-16), & x < 16, \\
x-16,     & x \geq 16
\end{cases}$ \\
Sigmoid & $f(x) = \frac{1}{1+\exp\left(-\frac{x-16}{2}\right)}$ \\
Exponential & $f(x) = \exp\!\left(\frac{x}{8}\right)$ \\
\midrule
\multicolumn{2}{@{}l}{\textit{Evaluation Metrics and Datasets}} \\
\midrule
Metrics & NRMSE = $\frac{\sqrt{MSE}}{f(32) - f(0)}$; throughput; energy; area\\
Datasets & Monte Carlo: 5000 searches per similarity level;  \\
& Distribution: 32/10000 random candidate/query vectors \\
\bottomrule
\end{tabularx}
\end{table}

\section{Evaluation}

RE-CAM enhances CAM array readout by reusing the CMP per ML and jointly optimizing the three design variables under system constraints and hardware nonidealities. This section evaluates whether the optimization algorithm improves approximation accuracy and examines the tradeoffs among different sensing circuits. The experimental setup is detailed below and summarized in TABLE~\ref{tab:setup}.

\subsection{Experimental Setup}

\textbf{Hardware Implementation:} Nominal ML voltage waveforms are SPICE-simulated using a $1\times32$ binary CAM array with 10T CMOS cells in 65 nm technology. All experiments use a $32\times32$ array. To balance throughput and area in the compact macro, ADC-based implementations use four 500-MS/s ADCs, each shared by eight MLs. All ML voltages are sampled and held at the selected timestamp. Including sensing circuitry, the array's energy and area are $11.92$ fJ/search/bit and $18{,}065\,\mu\mathrm{m}^{2}$ with U-ADC, and $1.48$ fJ/search/bit and $8{,}842\,\mu\mathrm{m}^{2}$ with CMP. For NU-ADC, we assume U-ADC's energy and area, optimistically excluding overhead for reconfigurable references. The reconfigurable decoder, simulated at the same node, supports the $32\times32$ array with FP16 decoding parameters, consuming $10.50$ pJ/query and occupying $5{,}339\,\mu\mathrm{m}^{2}$.

\textbf{System Constraints:} The CAM array supports a 500 MHz search frequency, with a 4 GHz clock for timing control unless otherwise specified. Since the simulated comparator latency $\tau_{\mathrm{cmp}}$ is 70 ps, shorter than the 250 ps clock cycle, timing control is the bottleneck for comparator reuse. This configuration allows up to 7 comparisons per ML per CAM search. We hence select 3-bit ADCs as our baselines, whose 7 references match this comparison budget.

\textbf{Hardware Nonidealities:} We set $\rho=10\%$ for ML discharge speed variation. For comparator residual offset and sensing noise, we use $\sigma_{\mathrm{off}}=\sigma_{\mathrm{n}}=1$ mV based on measurements reported in \cite{CmpNoise}.

\textbf{Baseline Implementation:} We evaluate RE-CAM against two baselines. U-ADC uniformly quantizes the ML voltage range and assigns each decoding category the mean target value of the similarity levels mapped to it under nominal waveforms. This baseline uses a conventional uniform-reference flash ADC \cite{FlashADC} with the RE-CAM decoder, providing flexibility in decoding parameters. NU-ADC uses $S+1$ evenly spaced decoding parameters spanning the function's output range, including its minimum and maximum, with nonuniform reference voltages selected accordingly. This baseline is inspired by NL-DPE \cite{NL-DPE}, RACE-IT \cite{RACE-IT}, and NL-ADC \cite{NL-ADC}, providing flexibility in reference voltages and decoding parameters when combined with the RE-CAM decoder. Both baselines select the sensing timestamp that minimizes approximation error.

\textbf{Target Functions:} We evaluate four representative functions: \textit{linear}, \textit{piecewise-linear}, \textit{sigmoid}, and \textit{exponential}, with the specific functions defined in TABLE~\ref{tab:setup}.

\textbf{Evaluation Metrics:} We measure accuracy using NRMSE, defined as root MSE divided by the function’s output range, with lower values representing higher accuracy. We also report search throughput, energy per search, and area to assess tradeoffs among sensing circuits.

\textbf{Datasets:} \textit{Monte Carlo} performs 5000 searches per similarity level to evaluate approximation accuracy across the full input range. \textit{Distribution} uses 32 candidate vectors and 10000 query vectors, all random 32-bit binary vectors, to evaluate accuracy under a nonuniform similarity distribution. Dataset-specific enhancement uses a separate training dataset for calibration.

\newcommand{\nrmsecell}[2]{\shortstack{#1\%\\(#2\%)}}

\begin{table*}[!t]
\centering
\caption{NRMSE (relative change compared with U-ADC), search throughput, energy per search, and area.}
\label{tab:comparison}
\footnotesize
\setlength{\tabcolsep}{2pt}
\renewcommand{\arraystretch}{1.12}
\begin{threeparttable}
\resizebox{\textwidth}{!}{%
\begin{tabular}{@{}l*{12}{c}@{}}
\toprule
& \multicolumn{6}{c}{\textbf{Monte Carlo}}
& \multicolumn{6}{c}{\textbf{Distribution}} \\
\cmidrule(lr){2-7}
\cmidrule(lr){8-13}
\textbf{Function / Implementation}
& U-ADC
& \shortstack{RE-CAM(U)}
& NU-ADC
& \shortstack{RE-CAM(N)}
& \shortstack{RE-CAM}
& \shortstack{Finer Sensing}
& U-ADC
& \shortstack{RE-CAM(U)}
& NU-ADC
& \shortstack{RE-CAM(N)}
& \shortstack{RE-CAM}
& \shortstack{Finer Sensing} \\
\midrule
Linear
& \nrmsecell{8.20}{0.00}
& \nrmsecell{7.92}{-3.38}
& \nrmsecell{7.52}{-8.28}
& \nrmsecell{6.67}{-18.66}
& \nrmsecell{6.90}{-15.82}
& \nrmsecell{6.43}{-21.50}
& \nrmsecell{8.52}{0.00}
& \nrmsecell{5.93}{-30.45}
& \nrmsecell{7.19}{-15.63}
& \nrmsecell{4.97}{-41.74}
& \nrmsecell{5.03}{-41.03}
& \nrmsecell{4.74}{-44.35} \\[3pt]
Piecewise-Linear
& \nrmsecell{7.16}{0.00}
& \nrmsecell{6.96}{-2.67}
& \nrmsecell{6.54}{-8.64}
& \nrmsecell{5.79}{-19.11}
& \nrmsecell{5.75}{-19.63}
& \nrmsecell{5.59}{-21.91}
& \nrmsecell{8.06}{0.00}
& \nrmsecell{5.62}{-30.35}
& \nrmsecell{6.69}{-17.07}
& \nrmsecell{4.88}{-39.45}
& \nrmsecell{4.91}{-39.11}
& \nrmsecell{4.66}{-42.25} \\[3pt]
Sigmoid
& \nrmsecell{12.64}{0.00}
& \nrmsecell{12.49}{-1.14}
& \nrmsecell{9.44}{-25.31}
& \nrmsecell{9.20}{-27.23}
& \nrmsecell{9.56}{-24.34}
& \nrmsecell{8.80}{-30.33}
& \nrmsecell{21.28}{0.00}
& \nrmsecell{17.52}{-17.63}
& \nrmsecell{16.52}{-22.35}
& \nrmsecell{14.69}{-30.94}
& \nrmsecell{15.01}{-29.43}
& \nrmsecell{14.22}{-33.18} \\[3pt]
Exponential
& \nrmsecell{4.62}{0.00}
& \nrmsecell{4.57}{-0.96}
& \nrmsecell{4.79}{+3.66}
& \nrmsecell{4.20}{-9.04}
& \nrmsecell{4.11}{-10.89}
& \nrmsecell{4.01}{-13.20}
& \nrmsecell{5.37}{0.00}
& \nrmsecell{3.27}{-39.08}
& \nrmsecell{5.16}{-3.76}
& \nrmsecell{2.90}{-45.93}
& \nrmsecell{2.86}{-46.73}
& \nrmsecell{2.72}{-49.39} \\
\midrule
Search Throughput (QPS)
& 62.5M & 62.5M & 62.5M
& 62.5M & 500M & 500M
& 62.5M & 62.5M & 62.5M
& 62.5M & 500M & 500M \\
Energy per Search (pJ)
& 22.71 & 22.71 & 22.71
& 22.71 & 12.02 & 12.02
& 22.71 & 22.71 & 22.71
& 22.71 & 12.02 & 12.02 \\
Area ($\mu\mathrm{m}^{2}$)
& 23,404 & 23,404 & 23,404
& 23,404 & 14,181 & 14,181
& 23,404 & 23,404 & 23,404
& 23,404 & 14,181 & 14,181 \\
\bottomrule
\end{tabular}%
}
\begin{tablenotes}[flushleft]
\scriptsize\raggedright
\setlength{\itemsep}{0pt}
\setlength{\parsep}{0pt}
\item[]
\end{tablenotes}
\end{threeparttable}
\end{table*}

\subsection{Performance Comparison and Ablation Study}

TABLE~\ref{tab:comparison} compares approximation accuracy, search throughput, energy per search, and area under Monte Carlo and Distribution datasets. RE-CAM(U) and RE-CAM(N) apply RE-CAM’s optimization framework to U-ADC and NU-ADC sensing, respectively. U-ADC and NU-ADC serve as baselines with the same RE-CAM decoder but baseline parameter settings. \textit{Finer Sensing} denotes RE-CAM with twice the clock rate and an unchanged comparison budget.

U-ADC uses uniform reference voltages at a single timestamp. By optimizing its decoding parameters, RE-CAM(U) reduces NRMSE by 0.96\%–3.38\% under Monte Carlo and 17.63\%–39.08\% under Distribution, demonstrating the benefit of noise-aware decoding and dataset-specific enhancement. NU-ADC adapts the reference voltages and decoding parameters to the target function, achieving higher accuracy than U-ADC in every evaluated case except the exponential function under Monte Carlo. The noticeable improvement supports our statement \textit{\textbf{approximating different functions prioritizes the separation of different similarity levels}}. The exponential result further supports this observation: uniformly quantizing the function’s output range concentrates reference voltages at higher similarity levels, leaving few to separate the many levels mapped to lower function values. RE-CAM(N) improves accuracy over NU-ADC in every evaluated case, demonstrating the benefit of gradient-based co-optimization over the rule-based configuration.

RE-CAM achieves approximation accuracy comparable to RE-CAM(N), with each outperforming the other on different target functions. However, by reusing one CMP per ML, RE-CAM reduces energy per search and area by 47\% and 39\%, respectively, presenting a more efficient alternative. We hypothesize that RE-CAM’s accuracy is constrained by the available sensing timestamps: some similarity levels may become difficult to separate before the next clock tick. To examine this limitation, we double the clock rate while retaining the same comparison budget and CAM search frequency. This configuration, denoted \textit{Finer Sensing}, expands the candidate timestamps from 7 to 15 while allowing at most 7 comparisons per CAM search. Finer Sensing improves accuracy over both RE-CAM and RE-CAM(N) for all evaluated functions under both datasets, supporting our hypothesis and the previous claim \textit{\textbf{different similarity levels are best separated with different sensing timestamps and reference voltages}}. We also evaluate doubling both the clock rate and CAM search frequency, which retains 7 candidate timestamps but shortens the search period from 2 ns to 1 ns. This configuration can, however, perform worse than the default configuration, suggesting that observations between 1 ns and 2 ns provide useful separation that earlier sensing cannot fully replace, strengthening the support for our claim.

Together, these results provide experimental support for the two observations motivating RE-CAM in the introduction, reinforcing the importance of jointly addressing \textit{when to sense, which reference to compare with, and how to decode the binary observations}.

RE-CAM overlaps sensing and decoding to retain native throughput across clock rates and comparison budgets, provided decoding keeps pace with CAM searches. At 4 GHz, its decoder processes 8 LUT entries in 2 ns, retaining 500M queries per second (QPS). 4 shared 500-MS/s ADCs require 8 conversion rounds per query, limiting ADC-based implementations to 62.5M QPS even with overlapped decoding. RE-CAM thus achieves $8\times$ higher search throughput.

\begin{table}[!t]
\centering
\caption{relationship between comparison budget and nrmse\\(relative change compared with a comparison budget of 3)}
\label{tab:sensing_budget}
\footnotesize
\begin{tabularx}{\columnwidth}
{@{}l*{3}{>{\centering\arraybackslash}X}@{}}
\toprule
\textbf{Dataset / Comparison Budget} & 3 & 7 & 15 \\
\midrule
Monte Carlo
& \shortstack{9.08\%\\(0.00\%)}
& \shortstack{6.58\%\\(-27.48\%)}
& \shortstack{5.60\%\\(-38.25\%)} \\
Distribution
& \shortstack{7.90\%\\(0.00\%)}
& \shortstack{6.95\%\\(-12.05\%)}
& \shortstack{6.54\%\\(-17.22\%)} \\
\bottomrule
\end{tabularx}
\end{table}

We further examine the effect of the comparison budget. TABLE~\ref{tab:sensing_budget} shows that increasing the budget from 3 to 7 and 15 under the same search throughput reduces the average NRMSE for both datasets. These results indicate that additional comparisons can improve accuracy by providing more observations and decoding categories.

Finer sensing and larger comparison budgets can be supported by increasing the clock rate or ML capacitance, using multiple CMPs per ML, or performing multiple CAM searches per query. These techniques trade accuracy against search throughput, energy, and area, making the configuration highly application-dependent. It is worth noting that we evaluate RE-CAM with a CMOS-based CAM array, which discharges faster than most alternative CAM technologies, including ReRAM, PCM, STT-MTJ \cite{SlowDischarge1}, FeFET \cite{SlowDischarge2}, and Flash \cite{SlowDischarge3}. Using those CAM technologies could enable finer sensing and more comparisons, both potentially leading to higher accuracy.

\section{Conclusion}
While IMS holds great potential in accelerating search-based data-intensive applications, binary CAM array readout limits application performance and broader applicability. Existing quantitative readout approaches either target Hamming-distance recovery or do not account for time-varying ML voltages, limiting their flexibility for CAM-based function approximation. This work presents RE-CAM, a readout enhancement framework for CAM that is agnostic to memory technology and application. RE-CAM reuses one CMP per ML for repeated sensing within a single CAM search and jointly optimizes sensing timestamps, reference voltages, and decoding parameters under system constraints and hardware nonidealities. By co-designing the optimization algorithm and reconfigurable hardware, RE-CAM improves approximation accuracy and reduces energy and area overhead while retaining the CAM array's native search throughput. These capabilities could improve application accuracy and broaden CAM adoption for tasks requiring quantitative similarity information.

\section{content generated by ai}
Codex assisted in code development and manuscript polishing, with all generated content reviewed and verified by the authors.

\bibliographystyle{IEEEtran}
\bibliography{references}

\end{document}